\documentclass{elsart}
\usepackage{epsfig}
\def\kms{\kern-1mm}
\def\ifm#1{\relax\ifmmode#1\else$#1$\fi}   
\def\dif{\hbox{d}}
\def\to{\ifm{\rightarrow}} 
\def\sig{\ifm{\sigma}}   
\def\plm{\ifm{\pm}}
\def\f{\ifm{\phi}}  
\def\epm{\ifm{e^+e^-}} 
\def\mupm{\ifm{\mu^+\mu^-}}
\def\ab{\ifm{\sim}} 
\def\x{\ifm{\times}}   
\def\deg{\ifm{^\circ}}
\def\up#1{$^{#1}$}  
     
\def\pic{\ifm{\pi^+\pi^-}}
\def\mev{{\rm \,Me\kern-0.125em V}}
\def\kev{{\rm \,ke\kern-0.125em V}}
\def\ev{{\rm \,e\kern-0.125em V}}
\def\baba{$~e^+e^-\to e^+e^-$}
\def\mumu{$~e^+e^-\to \mu^+\mu^-$}
\def\tot{$~e^+e^-\to e^+e^-,~\mu^+\mu^-$}
\catcode`@=11
\newdimen\z@ \z@=0pt
\newskip\z@skip \z@skip=0pt plus0pt minus0pt
\def\m@th{\mathsurround=\z@}
\def\ialign{\everycr{}\tabskip\z@skip\halign} 
\def\eqalign#1{\null\,\vcenter{\openup\jot\m@th
    \ialign{\strut\hfil$\displaystyle{##}$&$\displaystyle{{}##}$\hfil
      \crcr#1\crcr}}\,}
\newcommand{\aff}[2]{Dipartimento di Fisica dell'Universit\`a #1 e Sezione INFN, #2, Italy.}
\newcommand{\affd}[1]{Dipartimento di Fisica dell'Universit\`a e Sezione INFN, #1, Italy.}

\begin{document}
\begin{frontmatter}
\title{Measurement of the leptonic decay widths of the \f-meson with the KLOE detector}
\collab{The KLOE Collaboration}
\author[Na]{F.~Ambrosino},
\author[Frascati]{A.~Antonelli},
\author[Frascati]{M.~Antonelli},
\author[Roma3]{C.~Bacci},
\author[Roma3]{M.~Barva},
\author[Frascati]{P.~Beltrame},
\author[Frascati]{G.~Bencivenni},
\author[Frascati]{S.~Bertolucci},
\author[Roma1]{C.~Bini},
\author[Frascati]{C.~Bloise},
\author[Roma1]{V.~Bocci},
\author[Frascati]{F.~Bossi},
\author[Frascati,Virginia]{D.~Bowring},
\author[Roma3]{P.~Branchini},
\author[Moscow]{S.~A.~Bulychjov},
\author[Roma1]{R.~Caloi},
\author[Frascati]{P.~Campana},
\author[Frascati]{G.~Capon},
\author[Na]{T.~Capussela},
\author[Roma2]{G.~Carboni},
\author[Roma3]{F.~Ceradini},
\author[Pisa]{F.~Cervelli},
\author[Frascati]{S.~Chi},
\author[Na]{G.~Chiefari},
\author[Frascati]{P.~Ciambrone},
\author[Virginia]{S.~Conetti},
\author[Frascati]{E.~De~Lucia},
\author[Roma1]{A.~De~Santis},
\author[Frascati]{P.~De~Simone},
\author[Roma1]{G.~De~Zorzi},
\author[Frascati]{S.~Dell'Agnello},
\author[Karlsruhe]{A.~Denig},
\author[Roma1]{A.~Di~Domenico},
\author[Na]{C.~Di~Donato},
\author[Pisa]{S.~Di~Falco},
\author[Roma3]{B.~Di~Micco},
\author[Na]{A.~Doria},
\author[Frascati]{M.~Dreucci}
       \footnote{Corresponding author: M.~Dreucci, Laboratori Nazionali
	 di Frascati dell'INFN,via E.Fermi,40,00044 Frascati(RM),Italy.
	{\it~Email address Marco.Dreucci@lnf.infn.it}},
\author[Roma3]{A.~Farilla},
\author[Frascati]{G.~Felici},
\author[Karlsruhe]{A.~Ferrari},
\author[Frascati]{M.~L.~Ferrer},
\author[Frascati]{G.~Finocchiaro},
\author[Frascati]{C.~Forti},
\author[Roma1]{P.~Franzini},
\author[Roma1]{C.~Gatti},
\author[Roma1]{P.~Gauzzi},
\author[Frascati]{S.~Giovannella},
\author[Lecce]{E.~Gorini},
\author[Roma3]{E.~Graziani},
\author[Pisa]{M.~Incagli},
\author[Karlsruhe]{W.~Kluge},
\author[Moscow]{V.~Kulikov},
\author[Roma1]{F.~Lacava},
\author[Frascati]{G.~Lanfranchi},
\author[Frascati,StonyBrook]{J.~Lee-Franzini},
\author[Karlsruhe]{D.~Leone},
\author[Frascati,Moscow]{M.~Martemianov},
\author[Frascati]{M.~Martini},
\author[Na]{P.~Massarotti},
\author[Frascati,Moscow]{M.~Matsyuk},
\author[Frascati]{W.~Mei},
\author[Na]{S.~Meola},
\author[Roma2]{R.~Messi},
\author[Frascati]{S.~Miscetti},
\author[Frascati]{M.~Moulson},
\author[Karlsruhe]{S.~M\"uller},
\author[Frascati]{F.~Murtas},
\author[Na]{M.~Napolitano},
\author[Roma3]{F.~Nguyen},
\author[Frascati]{M.~Palutan},
\author[Roma1]{E.~Pasqualucci},
\author[Frascati]{L.~Passalacqua},
\author[Roma3]{A.~Passeri},
\author[Frascati,Energ]{V.~Patera},
\author[Na]{F.~Perfetto},
\author[Roma1]{L.~Pontecorvo},
\author[Lecce]{M.~Primavera},
\author[Frascati]{P.~Santangelo},
\author[Roma2]{E.~Santovetti},
\author[Na]{G.~Saracino},
\author[StonyBrook]{R.~D.~Schamberger},
\author[Frascati]{B.~Sciascia},
\author[Frascati,Energ]{A.~Sciubba},
\author[Pisa]{F.~Scuri},
\author[Frascati]{I.~Sfiligoi},
\author[Frascati,Novosib]{A.~Sibidanov},
\author[Frascati]{T.~Spadaro},
\author[Roma3]{E.~Spiriti},
\author[Frascati,Tiblisi]{M.~Tabidze},
\author[Roma1]{M.~Testa},
\author[Roma3]{L.~Tortora},
\author[Roma1]{P.~Valente},
\author[Karlsruhe]{B.~Valeriani},
\author[Frascati]{G.~Venanzoni},
\author[Roma1]{S.~Veneziano},
\author[Lecce]{A.~Ventura},
\author[Roma3]{R.~Versaci},
\author[Na]{I.~Villella},
\author[Frascati,Beijing]{G.~Xu}
\clearpage
\address[Beijing]{Permanent address: Institute of High Energy Physics of Academica Sinica, Beijing, China.}
\address[Frascati]{Laboratori Nazionali di Frascati dell'INFN, Frascati, Italy.}
\address[Karlsruhe]{Institut f\"ur Experimentelle Kernphysik, Universit\"at Karlsruhe, Germany.}
\address[Lecce]{\affd{Lecce}}
\address[Moscow]{Permanent address: Institute for Theoretical and Experimental Physics, Moscow, Russia.}
\address[Na]{Dipartimento di Scienze Fisiche dell'Universit\`a ``Federico II'' e Sezione INFN, Napoli, Italy}
\address[Novosib]{Permanent address: Budker Institute of Nuclear Physics, Novosibirsk, Russia.}
\address[Pisa]{\affd{Pisa}}
\address[Energ]{Permanent address: Dipartimento di Energetica dell'Universit\`a ``La Sapienza'', Roma, Italy.}
\address[Roma1]{\aff{``La Sapienza''}{Roma}}
\address[Roma2]{\aff{``Tor Vergata''}{Roma}}
\address[Roma3]{\aff{``Roma Tre''}{Roma}}
\address[StonyBrook]{Permanent address: Physics Department, State University of New York at Stony Brook, USA.}
\address[Tiblisi]{Permanent address: High Energy Physics Institute, Tbilisi State University, Tbilisi, Georgia.}
\address[Virginia]{Physics Department, University of Virginia, USA.}
\begin{abstract}  
The $\phi$-meson leptonic widths, $\Gamma_{ee}$ and $\Gamma_{\mu\mu}$, are obtained, respectively, from the \epm\ forward-backward asymmetry and the \mupm\ cross section around the \f-mass energy. We find $\Gamma_{ee}$=1.32\plm 0.05\plm 0.03\kev~and $\sqrt{\Gamma_{ee}\Gamma_{\mu\mu}}$=1.320\plm 0.018\plm 0.017\kev. These results, compatible with $\Gamma_{ee}$=$\Gamma_{\mu\mu}$, provide a precise test of lepton universality. Combining the two results gives $\Gamma_{\ell\ell}(\f)=1.320\pm0.023$ \kev.
\\
\vspace{.5cm}
{\it key words:}~leptonic width 
\end{abstract}
\end{frontmatter}
\section{Introduction}
\begin{figure}[ht]
  \centering
  \epsfig{file=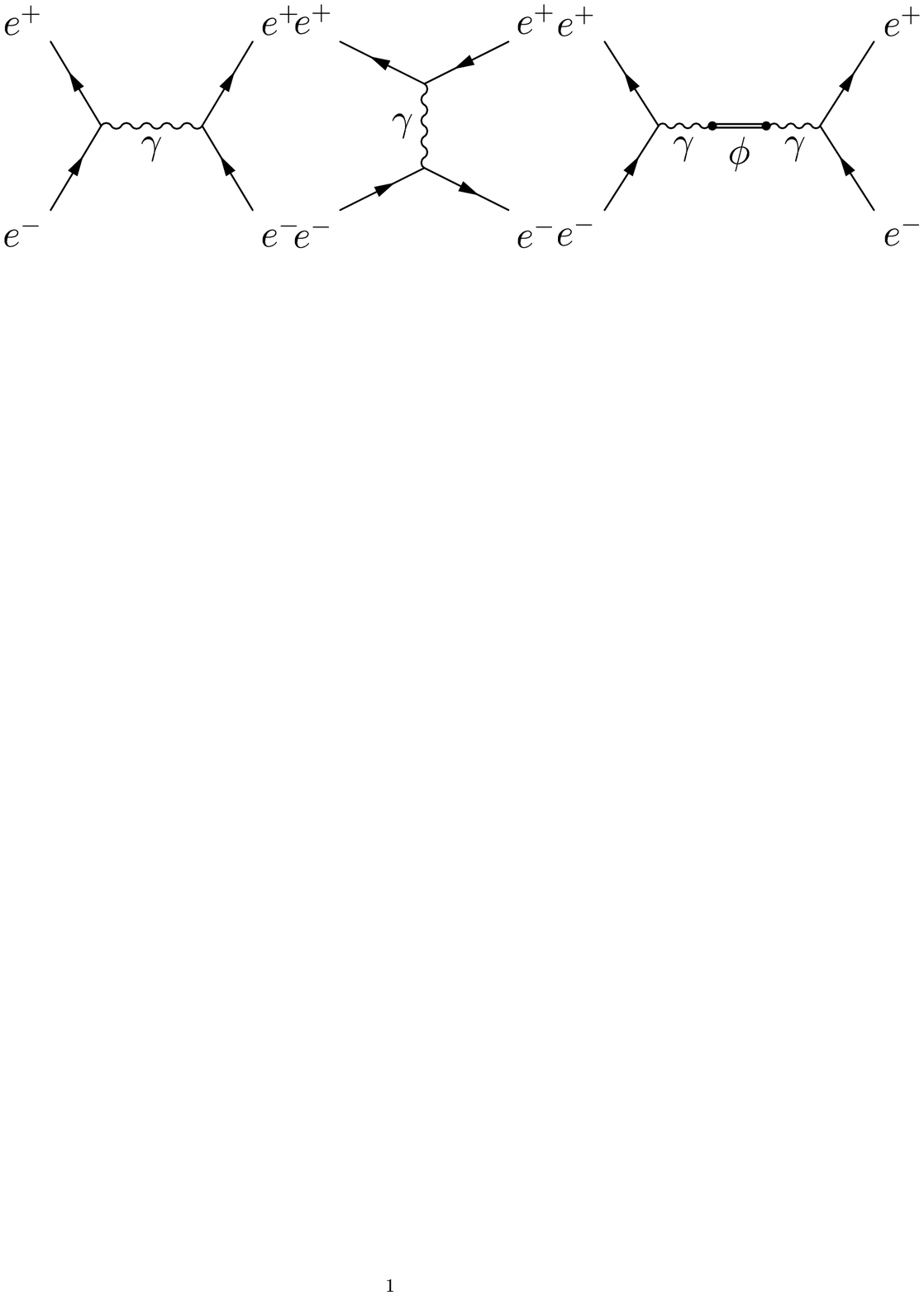,width=7cm}\kern1cm   \epsfig{file=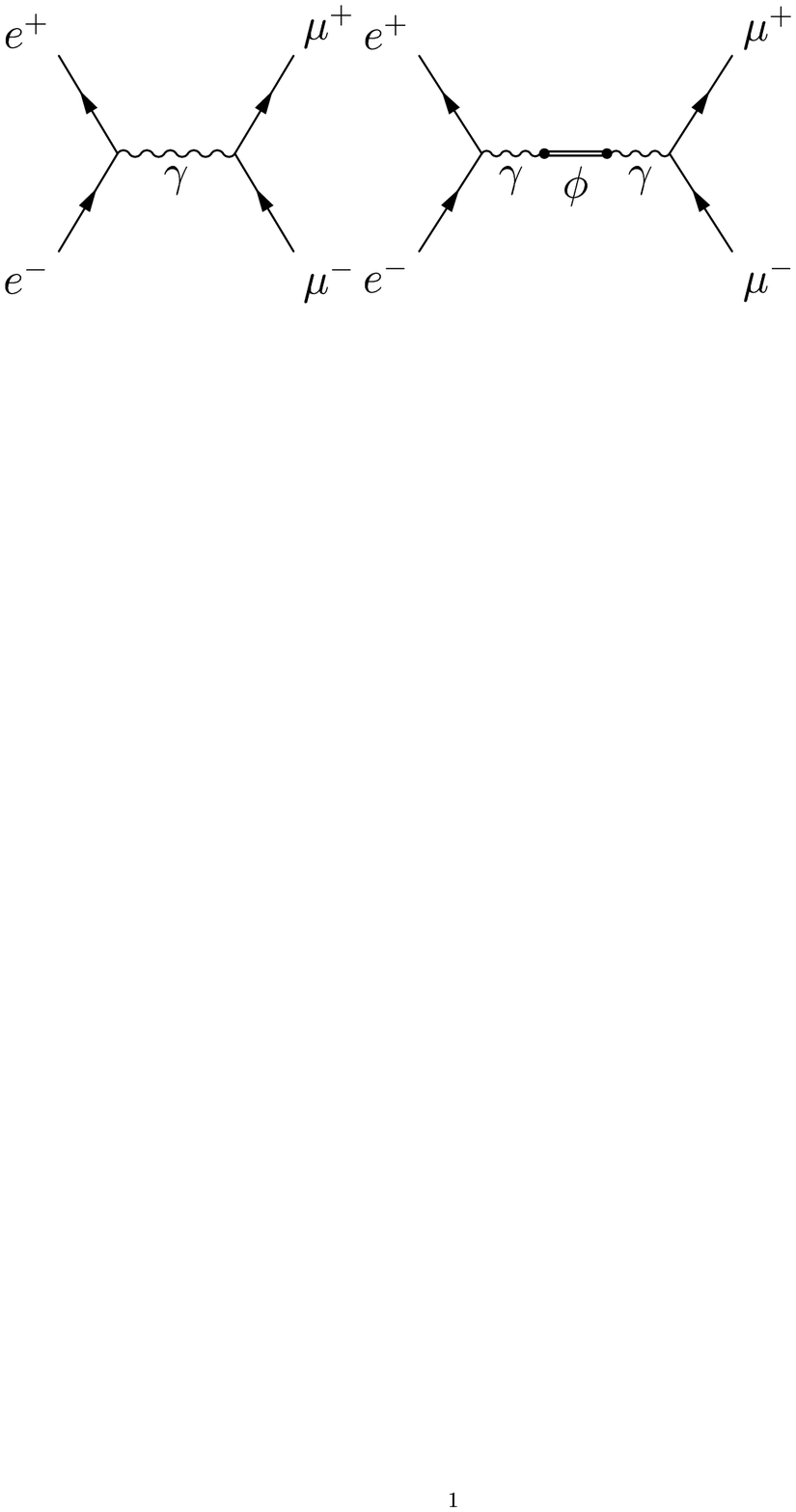,width=5cm} \caption {Feynman diagrams for \baba, left and \mumu, right, at $W\ab M(\f)$}  
  \label{fey1}
\end{figure}

\begin{figure}[ht]
  \centering
  \epsfig{file=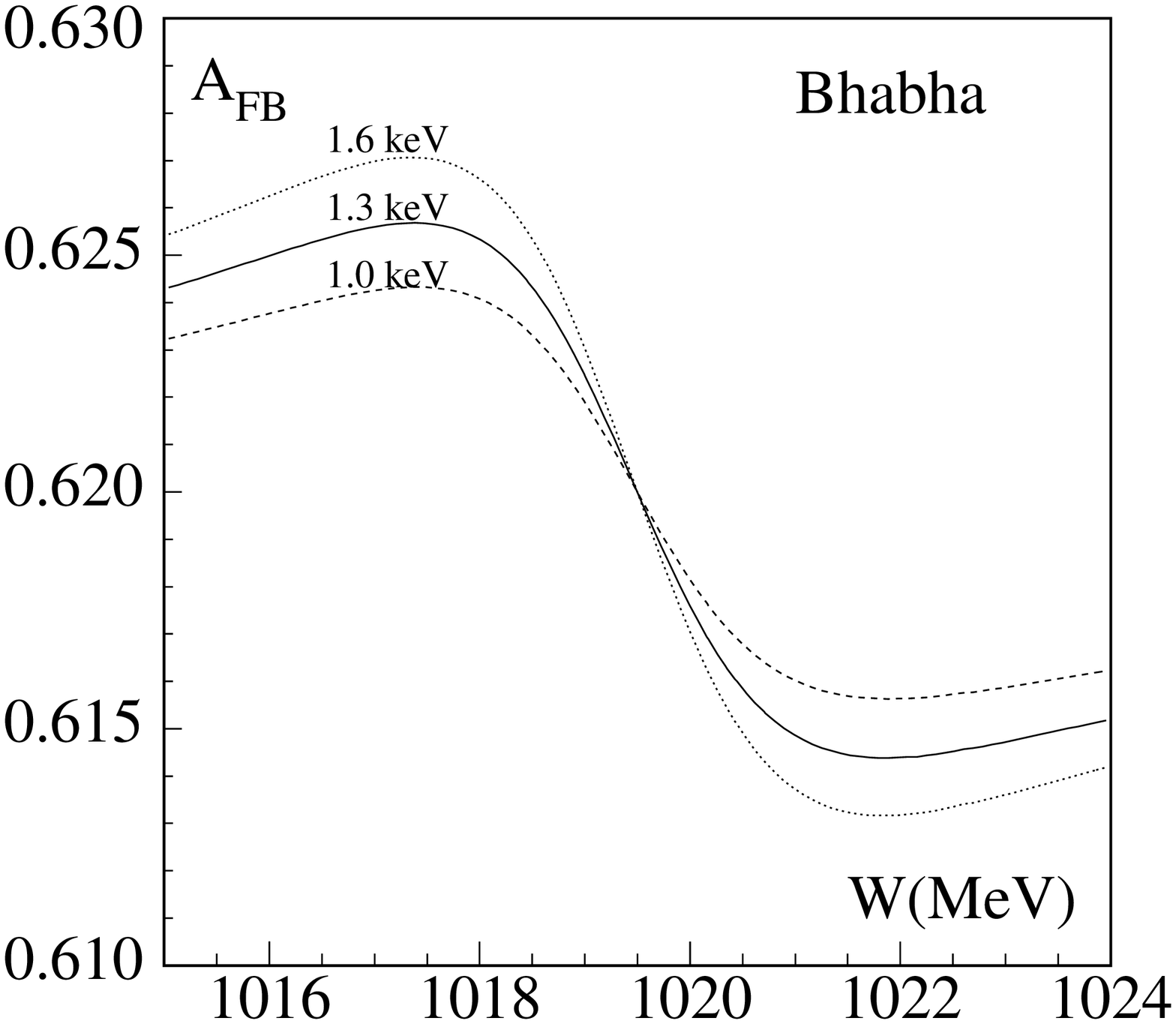,width=6cm}
  \epsfig{file=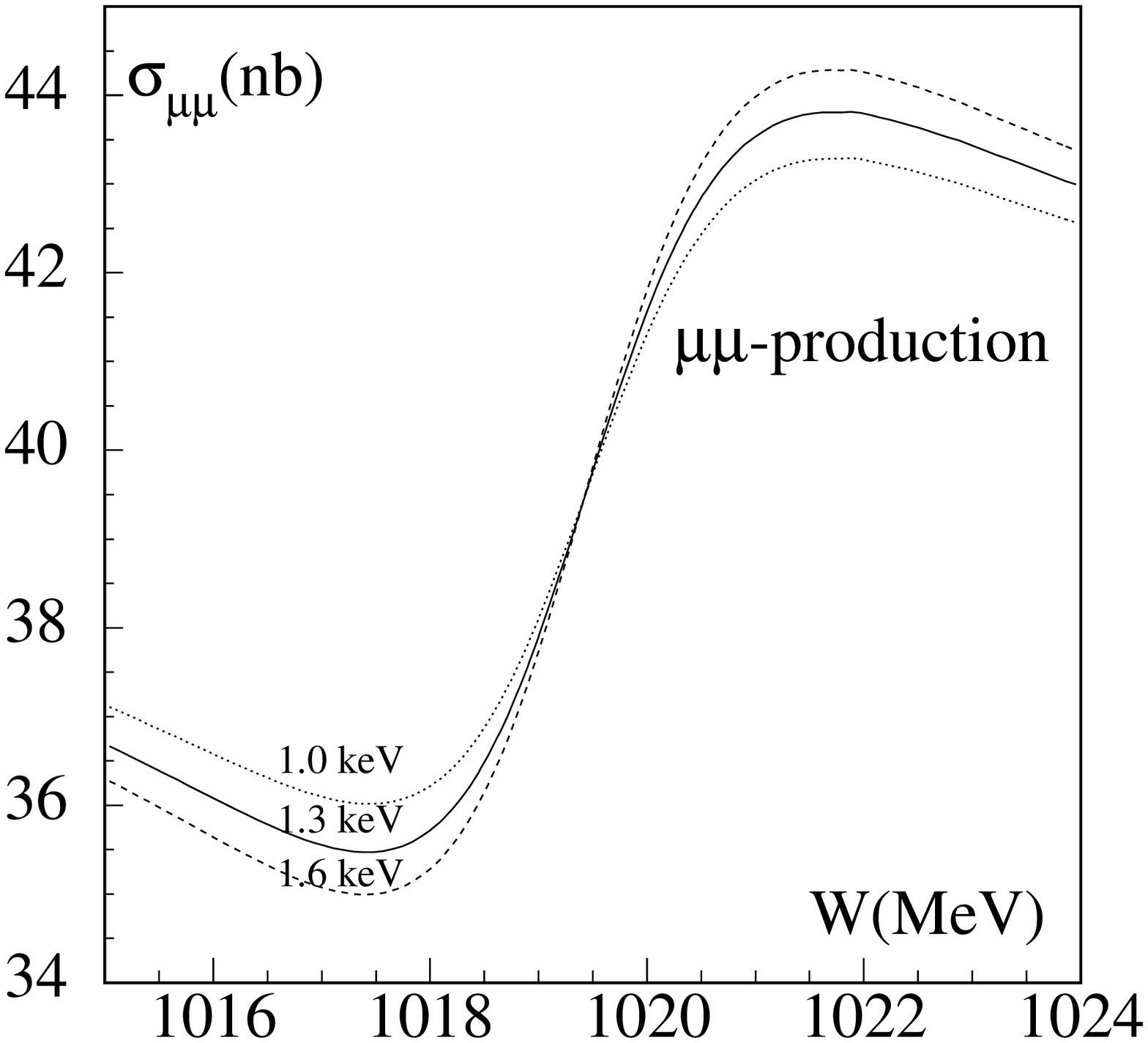,width=6cm}
  \caption{Left: $A_{FB}$ sensitivity to $\Gamma_{ee}$, for $\theta_0$=53\deg. Right: $\sigma_{\mu\mu}$ sensitivity to $\sqrt{\Gamma_{ee}\Gamma_{\mu\mu}}$, for $\theta_0$=50\deg. The three curves correspond to $\Gamma_{ll}$= 1.0, 1.3 and 1.6 \kev.} 
  \label{sensi1}       
\end{figure}

The \f-leptonic widths provide information on the \f-structure and its production cross section in \epm annihilations. They are necessary for decay branching ratio measurements and estimates of the hadronic contribution to vacuum polarization~\cite{ref1,ref2}. There is no direct measurement of the leptonic width. The only direct measurement is ~$\sqrt{B(\phi\to\epm)B(\phi\to\mupm)}$=(2.89\plm 0.10\plm 0.06)\x10\up{-4} ~\cite{acha1}. In Fig.~\ref{fey1} are shown the Feynman diagrams describing at the lowest order the processes\tot.
The cross sections can be written as $\sigma_{ee} = \sigma_{ee,\phi}+ \sigma_{\rm int}$ and $\sigma_{\mu\mu} = \sigma_{\mu\mu,\phi} +\sigma_{\rm int}$, with $\sigma_{\rm int}$, the interference term, given by:
\begin{equation}
  \sigma_{\rm int} =\frac{3\alpha\Gamma_{\ell\ell}}{M_\phi}\:
  \frac{W^2-M_\phi^2}{(W^2-M_\phi^2)^2+W^2\Gamma_\phi^2}\:
  \int^{\cos\theta_{max}}_{\cos\theta_{min}} f_{\ell\ell}(\theta)~{\rm d}\cos\theta.
  \label{interfer}
\end{equation}
where W is the energy in the collision center of mass (CM), $\theta_{min}$ and $\theta_{max}$ define the acceptance in the polar angle $\theta$ (see later) and where $\Gamma_{\ell\ell}=\Gamma_{ee}$ for \baba\ and $\Gamma_{\ell\ell}=\sqrt{\Gamma_{ee}\Gamma_{\mu\mu}/\xi}$ for \mumu. The $\xi$ term takes into account for the phase space correction :
\begin{equation}
  \xi=(1+2\frac{m^2_\mu}{M^2_\phi})(1-4\frac{m^2_\mu}{M^2_\phi})^{1/2}
\end{equation}
corresponding to 0.9993 with a negligible error. The interference term $\sigma_{\rm int}$ is linear in $\Gamma_{\ell\ell}$ and it changes sign when W goes through the pole at $M_{\phi}$. The angular distribution $f_{\ell\ell}(\theta)$ is:
\begin{equation}
  \eqalign{
    f_{ee}(\theta)&=\pi\left[1+\cos^2\theta-\frac{(1+\cos\theta)^2} {1-\cos\theta}\right]\cr
    f_{\mu\mu}(\theta)&=\pi\beta_\mu\left[1+\cos^2\theta+ (1-\beta_\mu^2)\sin^2\theta\right]\cr}
\end{equation}
with $\beta_{\mu}$ the muon velocity. In the following we use data collected at CM energies $W$ of 1017.2 and 1022.2 \mev, i.e. $M_{\phi}\pm\Gamma_\phi/2$, and at the $\phi$ peak, $W$=1019.7 \mev. For \mumu\ we measure the cross section. Since Bhabha scattering is dominated by the photon exchange amplitude, the interference term is best studied in the forward-backward asymmetry, $A_{FB}$, defined as
\begin{equation}
  A_{FB}= \frac{\sigma_F-\sigma_B}{\sigma_F+\sigma_B}.
  \label{asy}
\end{equation}
where $\sigma_F$ and $\sigma_B$ are the cross sections for events with electrons in the forward and backward hemispheres. Because of the strong divergence of the cross section at small angle, it is better to use an angular region around 90\deg, $\theta_0<\theta<180-\theta_0$. We use $\theta_0$=53\deg (50\deg) for \baba (\mumu). Fig.~\ref{sensi1} shows $A_{FB}$ and $\sigma_{\mu\mu}$ vs $W$ for $\Gamma_{\ell\ell}$= 1.0, 1.3 and 1.6 \kev. A 10\% change of $\Gamma_{\ell\ell}$  results in a fractional change of $\sim 10^{-3}$ and $\sim 4\x10^{-3}$ for $A_{FB}$ and $\sigma_{\mu\mu}$.
\\Radiative corrections for initial state radiation (ISR) and final state radiation (FSR) are included following \cite{teoria1,teoria2}. ISR, where one or both the colliding particles radiate a photon before interacting, reduces the CM energy, from the nominal value, $W$, to a lower value, $W'$. ISR photon(s), emitted mostly collinear to the beam, are not detectable. ISR-FSR interference effect results in an forward-backward asymmetry. To enhance the $\phi$-meson contribution, a lower cut on $W'/W$ is imposed for both \baba\ and \mumu\ processes. We define the variable:
\begin{equation}
  z = \sqrt{\frac{\sin\theta_++\sin\theta_--|\sin(\theta_++\theta_-)|}
    {\sin\theta_++\sin\theta_-+|\sin(\theta_++\theta_-)|}}             
  \label{sps}
\end{equation} 
where $\theta_+$ and $\theta_-$ are the angles of the final leptons in the $e^+e^-$ collision CM respect to the beam direction. In the hypothesis of a single collinear ISR photon and for FSR photon(s) collinear with the final leptons we have $W'/W=z$. If an event does not satisfy this assumption this relation is no more exact. To define our geometrical acceptance we use the polar angle in the CM of the final leptons. So in the following we refer to the polar angle $\theta$ as the average of the angles $\theta_-$ and ($180\deg-\theta_+$) in this frame.

The KLOE detector~\cite{geanfi} consists of a large volume drift chamber, DC, surrounded by an electromagnetic calorimeter, ECAL. They are immersed in a solenoidal magnetic field of about 0.52 T. For electron and muon tracks in the angular region defined above the momentum resolution is about 3\x10\up{-3}, while the angular resolution is $\sim$ 3 mrad. Vertexes inside the DC are reconstructed with a spatial resolution of about 3mm. The calorimeter, consisting of a barrel covering $45\deg<\theta<135\deg$ and two end caps, measures energy deposits with a resolution $\sigma(E)/E=5.7\%/\sqrt{E({\rm GeV})}$ and arrival times with a resolution $\sigma(t)=54/\sqrt{E({\rm GeV})}\oplus50$ ps. Large angle Bhabha scattering and $e^+e^-$ $\to \gamma\gamma$ events are used to measure luminosity, ${\mathcal L}$, and the beams crossing angle. The beam energy spread amounts to $\sim$ 330~\kev~for the CM energy.  The luminosity has an energy independent systematic uncertainty of 0.6\%.  The absolute energy scale has been established from a fine energy scan at the $\phi$ peak corresponding to \ab500 nb$^{-1}$ of integrated luminosity, using the CMD-2 value for the \f-meson mass: $M$(\f)=1019.483\plm0.027 \mev \cite{cmd2}. 
The same data have been used to perform a precision measurement of the neutral kaon mass. We find $M(K_S)$=497.583\plm0.005\plm0.020 $\rm MeV/c^2$ \cite{kaonmass}.

The CM energies and integrated luminosities ($\int\kern-1mm L\dif t$) for the data used in this analysis are summarized in Table~\ref{sample}. 
\begin{table}[hbt]
  \begin{center}
    \begin{tabular}{|c|c|}
      \hline
      \noalign{\vglue2pt}
      CM energy (\mev)   & $\int\kern-1mm L{\rm d}t$ (nb\up{-1}) \\
      \hline 
      $1017.17 \pm 0.01$ & $6966 \pm 42$ \\
      $1019.72 \pm 0.02$ & $4533 \pm 27$ \\
      $1022.17 \pm 0.01$ & $5912 \pm 35$ \\
      \hline
    \end{tabular}
    \caption{2002 $\phi$-scan statistics}
    \label{sample}
  \end{center}
\end{table}
Analysis efficiencies and resolutions are determined with the KLOE Monte Carlo (MC) program, in which different generators for \epm and \mupm are used~\cite{geanfi,babayaga}. Acceptance cuts and background are studied both with data and MC.

\section{$\Gamma_{ee}$ measurement }\label{sec:bha}
\begin{figure}[ht]
  \centering
  \epsfig{file=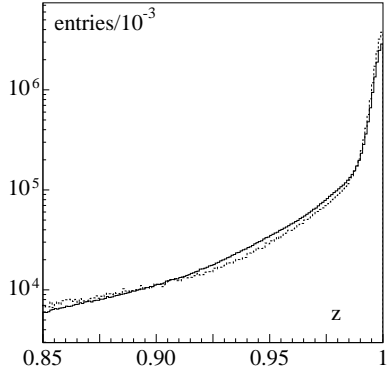,width=5.cm}
  \caption {\baba. $z$ reconstructed distribution for data (solid) and MC (dot)} 
  \label{mc1}  
\end{figure}

\begin{figure}[ht]
  \centering
  \epsfig{file=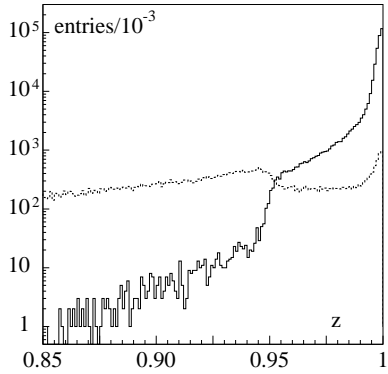,width=5.cm}
  \caption {\baba. $z$ reconstructed distribution for MC events: signal ($W'/W>0.95$, continuous) and radiative background ($W'/W<0.95$, dashed)} 
  \label{mc2}  
\end{figure}

Bhabha events are selected requiring at least two clusters with energy $E_{cl}$ in the range 300-800 \mev, a narrow time window, a total energy deposited in ECAL greater than 800 \mev, a cut on the angle between the two most
energetic clusters, $|180-\theta_1-\theta_2|<10^\circ$ and a polar angle acceptance $\sim 30^\circ<\theta<150^\circ$. We only use events for which the polar angle satisfies $53^\circ<\theta<127^\circ$, corresponding to the central region of the barrel calorimeter. Muon and pion contamination is quite negligible. For each event we evaluate $z$ as defined in eq.~\ref{sps}, and retain events with $z>0.95$. In Fig.~\ref{mc1} the distribution for reconstructed $z$ for a data sample and MC is shown. Fig.~\ref{mc2} shows the distribution of reconstructed $z$ for the signal ($W'/W>0.95$) and the radiative background ($W'/W<0.95$), obtained with MC. 
The efficiency and the radiative background have been evaluated  as a function of the polar angle. On average, an efficiency of about 98\% and a contamination of radiative background of about 2\% are obtained for $z>0.95$. 
In this analysis some of the systematics are independent of $W$ and do not affect the measurement of $\Gamma_{ee}$: they amount $\sim 0.2\%$ and are dominated by the uncertainty on FSR correction (0.15\% as evaluated by MC) and by the $\theta$ resolution (0.07\%). This resolution introduces no bias but reduces the asymmetry. We find the reduction to be $0.0011\pm 0.0004$, the same for the three energies. Note that the asymmetry measurement does not depend on luminosity and trigger efficiency. Systematic errors on $A_{FB}$ due to $z$ resolution, $\theta$ cut and background estimate are evaluated by varying the acceptance cuts. 
They are summarized in Table~\ref{babasyst}.
\begin{table}[hbt]
  \centering
  \begin{tabular}{|c||c|c|c|}
    \hline       
    Fiducial cuts & $W_1$  & $W_2$  & $W_3$ \\
    \hline
    $0.90\!<\!z\!<\!0.98$&0.8&0.3&1.1\\
    $50\deg\!<\!\theta_0\!<\!70\deg$&1.0&1.0&1.0\\
    \hline
  \end{tabular}  
  \caption{Absolute systematic uncertainties on $A_{FB}$ in units of 10\up{-4}. $W_1$, $W_2$ and $W_3$ are respectively 1017.17, 1019.72 and 1022.17 \mev.}
  \label{babasyst}    
\end{table}

Table~\ref{asidata} shows the measured asymmetry for $z>0.95$ and $53\deg<\theta<127\deg$, together with the errors. The common systematic error of $\sim 0.2\%$ mentioned before is not included.
\begin{table}[hbt]
  \centering
  \begin{tabular}{|c|c|}
    \hline
    W (\mev)&  $A_{FB}$ \\
    \hline 
    1017.17 &0.6275\plm0.0003\\
    1019.72 &0.6205\plm0.0003\\
    1022.17 &0.6161\plm0.0004\\
    \hline
  \end{tabular}
  \caption{Forward-backward asymmetry results for $z>0.95$ and $53\deg<\theta<127\deg$}
  \label{asidata}
\end{table}
To fit these results we first convolute the cross section with the radiator function \cite{teoria1,teoria2} in order to account for ISR. We then fold-in the beam energy spread. 
The $\omega$-exchange contribution is expected to be very small, because $m_\phi-m_\omega\ab29\x\Gamma_\omega$. In fact, its inclusion in the amplitude produces at most a variation on $\Gamma_{ee}$ of less than 0.1\%. We have also verified that the $\rho$-exchange contribution is at the same level. The fit parameters are the leptonic width, $\Gamma_{ee}$, the $\phi$ mass, $M_{\phi}$, and the forward-backward asymmetry at the $\phi$ mass, $A^0$ . The total \f\ width used in eq.~\ref{interfer}, $\Gamma_{\phi}=4.26\pm0.05$ \mev, is taken from PDG~\cite{pdg}, giving an uncertainty of 0.013 \kev~for $\Gamma_{ee}$. 
The systematic error on $M_{\phi}$ stems from the determination of $W$,~\ab30~\kev~\cite{pdg}. The results of the fit to the data are shown in Table~\ref{geeresu}. 
\begin{table}[hbt]
  \begin{center}
    \begin{tabular}{|c|c|c|c|}
      \hline
      \kms Parameter\kms      &  value  & error & error \\[-3mm]
      &    & (stat) & (syst) \\
      \hline 
      \kms$\Gamma_{ee}$ (\kev)\kms& 1.32    & 0.05       & 0.03   \\
      $A^0$                      & 0.6215  & 0.0002     & 0.002  \\
      \kms$M_{\phi}$ (\mev)\kms&\kms1019.50\kms& 0.08       & 0.03   \\
      \hline
    \end{tabular}  
    \caption{Fit results for~\baba~process}
    \label{geeresu}
  \end{center}
\end{table}

\def\uu{\vphantom{\hbox{I}}}

\section{\mathversion{bold}$\sqrt{\Gamma_{ee}\Gamma_{\mu\mu}}$ measurement}

\begin{figure}[htb]
  \centering
  \epsfig{file=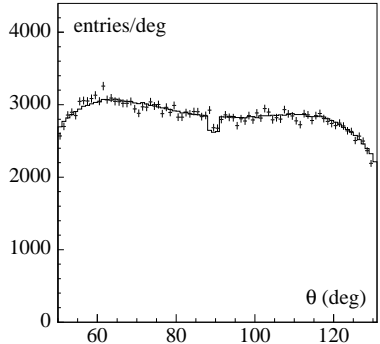,width=5cm}
  \caption{$\theta$ distribution. Data and MC simulation are in fair agreement. } \label{theta_asy}
\end{figure}

Identification of \mumu\ events is primarily based on the ratio $p/E_{\rm ECAL}$ between momentum measured in the DC and energy measured in ECAL in order to reject the large signal from Bhabha scattering. We require two tracks with $970<|\vec p_+|+ |\vec p_-|<1010$ \mev\ and a total energy signal from ECAL of less than 700 \mev. We accept the angular interval $50\deg<\theta<130\deg$. The residual background is mostly due to $ee\to\pi\pi$ and is about half of the $\mu$-pair events. 
The angular distribution distortion with respect to $1+\cos^2\theta$ due to ISR-FSR interference, is shown in Fig.~\ref{theta_asy}, showing a comparison between MC and an almost pure sample of \mumu events. 
Just as for Bhabha scattering, we require $z>0.985$. $\theta$ and $z$ are computed as before, sec. \ref{sec:bha}.

\begin{figure}[ht]
  \centering
  \epsfig{file=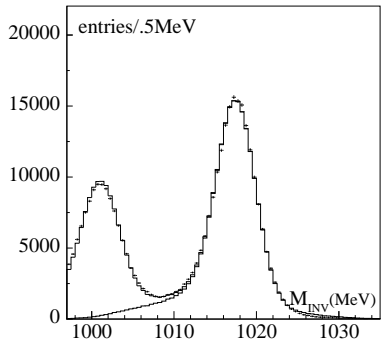,width=5cm}
  \caption {Fit to the invariant mass, computed assuming $m=m_\mu$, distribution of \pic\ and \mupm\ events for $W$=1017 \mev.} 
  \label{counting}
\end{figure}

Muon (signal) and pion (background) counting is obtained from fitting the two particle (assumed to be muons) invariant mass distributions to a the respective MC predictions, for each collider energy. A fit example is shown in Fig.~\ref{counting}. The invariant mass resolution and the beam energy spread are properly disentangled using data. 

Efficiencies due to the trigger and off-line filters have been obtained from data, by reprocessing a sample of unbiased raw data. All the above uncertainties are energy independent and affect the cross-section measurement but not $\Gamma_{\ell\ell}$. The fractional uncertainty for trigger and filters\ efficiencies is 0.7\%, for tracking efficiency is 0.5\% and for the ECAL energy cut is 0.5\% for a total of 1.0\%.
Including also the luminosity uncertainty (0.6\%) gives a total energy independent systematic error of 1.2\% for $\sigma_{\mu\mu}$. The systematic error due to background ($ee\to\pi\pi$) counting is \ab0.0045 nb. The uncertainties due to the $z$ and $\theta$ cuts are 0.01 nb, and 0.002 nb, respectively. The Bhabha scattering contamination is negligible, $\ll0.2\%$.
Table~\ref{xsmu} gives the cross section for \mumu.
\begin{table}[htb]
  \centering
  \begin{tabular}{|c|c|}
    \hline
    $W$ (\mev)&$\sigma_{\mu\mu}$ (nb)\\
    \hline 
    1017.17 & $35.66 \pm 0.08 $ \\
    1019.72 & $40.19 \pm 0.14 $ \\
    1022.17 & $43.92 \pm 0.09 $ \\
    \hline
  \end{tabular}
  \caption{Measured cross-section for $ee\to\mu\mu$. The energy independent errors discussed in the text are not included.}
  \label{xsmu}
\end{table}
To fit the data, the cross section is corrected for ISR and convoluted with the machine energy spread. Free parameters are the leptonic width, $\Gamma_{\ell\ell}=\sqrt{\Gamma_{ee}\Gamma_{\mu\mu}}$, the $\phi$ mass, $M_{\phi}$, and the $\mu\mu$ cross section at the $\phi$ mass, $\sigma_0$.

The \f\ width uncertainty adds an error of 0.013 \kev~to $\Gamma_{\ell\ell}$. The $\omega$ and $\rho$ exchange contribution is negligible. The FSR correction, evaluated by MC simulation, produces a 0.004 \kev~error  $\Gamma_{\ell\ell}$.
The result of the fit, including the systematics uncertainties listed above is:
$$\sqrt{\Gamma_{ee}\Gamma_{\mu\mu}}=1.320\plm0.018\plm0.017\kev$$
The fit returns a \f-meson mass of 1019.63\plm0.05 \mev and a cross section $\sig_0=39.2\pm0.04_{\rm stat}\pm0.4_{\rm syst}$~nb, being the expected one 39.6 nb. Both values confirm the validity of the measurement.

\section{Conclusion}
Our results, $\Gamma_{ee}=1.32\pm0.05\pm0.03$\kev, $\sqrt{\Gamma_{ee}\Gamma_{\mu\mu}}=1.320\pm0.018\pm0.017$\kev, are consistent with lepton universality. Accepting lepton universality for which there is ample evidence, combining the results we obtain: 
$$\Gamma_{\ell\ell}=1.320\plm0.017_{\rm stat}\plm0.015\,_{\rm syst}=1.320\pm0.023\ \hbox{\rm \kev}$$
The above result is a direct measurement of the leptonic width. It represents a considerable precision improvement over the only competing measurement $\sqrt{B(\phi\to\epm)B(\phi\to\mupm)}$ = (2.89\plm 0.10\plm 0.06)\x10\up{-4} \cite{acha1}.

\section{Acknowledgments}
We thank the DA$\Phi$NE team for their efforts in maintaining low
background running conditions and their collaboration during all
data-taking. We also thank F. Fortugno for his efforts in ensuring
good operations of the KLOE computing facilities. This work was
supported in part by DOE grant DE-FG-02-97ER41027; by 
EURODAPHNE, contract FMRX-CT98-0169; by the German Federal Ministry
of Education and Research (BMBF) contract 06-KA-957; 
by Graduiertenkolleg 'H.E. Phys.and Part. Astrophys.' of 
Deutsche Forschungsgemeinschaft, Contract No. GK 742;
by INTAS, contracts 96-624, 99-37; and by TARI, contract HPRI-CT-1999-00088.

\end{document}